\documentclass[aps,prl,twocolumn,superscriptaddress,floats,floatfix,preprintnumbers]{revtex4-1}

\usepackage{amssymb,amsmath}
\usepackage{bm,url,graphics,subfigure,epsfig}
\usepackage{ulem}
\usepackage{color}
\usepackage{soul}

\def \FF {F}
\def \igdot {\gdot^{-1}}
\def \Rparl {R_{\parallel}}
\def \rparl {r_{\parallel}}
\def \dz {d_{\rm z}}
\def \dy {d_{\rm y}}
\def \Sz {S^{\rm z}_2}
\def \zc {z_{\rm c}}
\def \rhop {\rho_{\rm p}}
\def \Up {{\bm U}^{\rm p}}
\def \Op {{\bm \Omega}^{\rm p}}
\def \IIp {\mathcal{I}_{\rm p}}
\def \EE {\bm{E}}
\def \taub {\bm{\tau}}
\def \Xl {\bm{X}_{\rm l}}
\def \Vp {V_{\rm p}}
\def \Uc {{\bm U}}
\def \uu {{\bm u}}
\def \nn {\bm{n}}
\def \rr {\bm{r}}
\def \ff {{\bm f}}

\def \Uzero {U_0}

\def \a {a}

\def \Lx {L_x}
\def \Lz {L_z}
\def \Ly {L_y}
\def \gdot {\dot{\gamma}}
\def \mue {\mu_{\rm eff}}
\def \muf {\mu_{\rm f}}

\def \xx {\bm{x}}

\def \grad {{\bm \nabla}}

\def \dive {{\bm \nabla}\cdot}
\def \lap {\nabla^2}
\def \delt {\partial_t}

\newcommand{\ddt}[1]{\frac{d #1}{dt}}

\newcommand{\bra}[1]{\langle #1\rangle}
{}

\def \rzero {r_{0}}
\def \Rey  {\mbox{Re}}

\def \EE{\bm E}

\newcommand{\eq}[1]{(\ref{#1})}

\newcommand{\fig}[1]{Fig.~(\ref{#1})}
\newcommand{\subfig}[2]{Fig.~(\ref{#1}{#2})}

\def \jfm {J. Flu. Mech.}

\definecolor{cinnamon}{rgb}{0.82, 0.41, 0.12}

\def \prl {Phys. Rev. Lett.}
\def \epl {Europhys. Lett.}

\def \ARFM {Annu. Rev. Flu. Mech.}

\begin{document}
\title{Rheology of confined non-Brownian suspensions}
\author{Walter~Fornari}
\affiliation{Linn\'e Flow Centre and SeRC, KTH Mechanics, SE-100 44 Stockholm, Sweden}
\author{Luca~Brandt}
\affiliation{Linn\'e Flow Centre and SeRC, KTH Mechanics, SE-100 44 Stockholm, Sweden}
\author{Pinaki Chaudhuri}
\affiliation{Institute of Mathematical Sciences, CIT Campus, Taramani, Chennai 600113, India}
\author{Cyan~Umbert~Lopez}
\affiliation{Linn\'e Flow Centre and SeRC, KTH Mechanics, SE-100 44 Stockholm, Sweden}
\author{Dhrubaditya~Mitra}
\affiliation{Nordita, KTH Royal Institute of Technology and Stockholm University,
Roslagstullsbacken 23, 10691 Stockholm, Sweden}
\author{Francesco~Picano}
\affiliation{Department of Industrial Engineering, University of Padova, Via Venezia 1, 35131, Padova, Italy}
\date{\today,~  }
\begin{abstract}
We study the rheology of confined suspensions of  neutrally buoyant rigid monodisperse spheres in plane-Couette
flow using Direct Numerical Simulations. 
We find that if the width of the channel is a (small) integer multiple of the sphere's
diameter, the spheres self-organize into two-dimensional layers
that slide on each other and the suspension's effective viscosity  is
significantly reduced.  Each two-dimensional layer is found to be structurally
liquid-like but their dynamics is frozen in time.
\end{abstract}
\preprint{NORDITA-2015-69}
\maketitle


Suspensions of solid objects in simple Newtonian solvents (e.g., water) can
show a kaleidoscope of rheological behaviors depending on the 
shape, size, volume fraction ($\phi$)  of the additives, and the shear-rate  ($\gdot$) imposed on the 
flow; see, e.g., Refs.~\cite{wag+bra09,sti+pow05} for reviews.
Suspensions can be of various types, e.g., 
suspensions of small particles (smaller than the viscous scale of the solvent), 
where the solvent plays the role of a thermal bath, are called
Brownian suspensions (e.g. colloids)~\cite{wag+bra09}. 
At small $\phi$ and under small $\gdot$, the effective viscosity of such suspensions increases
with $\phi$: $\mue \sim (5/2)\phi$ ~\cite{Bat67}, as derived by Einstein~\cite{ein06}
(See also Ref. \cite{bra84} for a $d$ dimensional generalization).
In the other category are suspensions of large particles (e.g.
emulsion, granular fluids etc.), where the thermal fluctuations are often
negligible.  Such suspensions are  called {\it non-Brownian} suspensions.
For moderate values of $\phi$ and $\gdot$  non-Brownian suspensions 
show {\it continuous shear-thickening}, (i.e.,  $\mue$ increases with $\gdot$)
 ~\cite{fal+lem+ber+bon+ova10,pic+bre+mit+bra13}
which can be understood~\cite{pic+bre+mit+bra13} in terms of excluded volume effects.
Such a rheological response has been observed in many natural and industrial flows, including flows of mud, lava, 
cement, and corn-starch solutions.  
Dense (large $\phi$, close to  random close packing) 
non-Brownian suspensions  may show  {\it discontinuous shear-thickening}~\cite{bro+jae09,isa}
-- a jump in effective viscosity as a function of $\gdot$. 

Recent experiments have uncovered intriguing rheological behavior of 
very dense (large $\phi$) suspensions under confinement~\cite{fal+lem+ber+bon+ova10, bro+jae09, brown+jaeger2010}. 
Common wisdom dictates that, under confinement, the inertial effects are generally 
unimportant at small $\gdot$. 
But a series of recent studies~\cite{dicar09,lee+ami+sto+dicar10,dicar+edd+hum+sto+ton09,
ami+sol+wea+dicar12} have demonstrated that the effect of fluid inertia, although small, can give rise
to variety of effects even in microfluidic flows. 
Drawing inspirations from these two sets of works, in this paper, we study the effects 
of confinement on non-Brownian suspensions with moderate values of $\phi$ and $\gdot$. 
In particular, we choose the range of $\phi$ and $\gdot$ such that in bulk the 
suspensions show continuous shear-thickening. 

\begin{figure*}[]
\begin{center}
\includegraphics[width=0.25\linewidth]{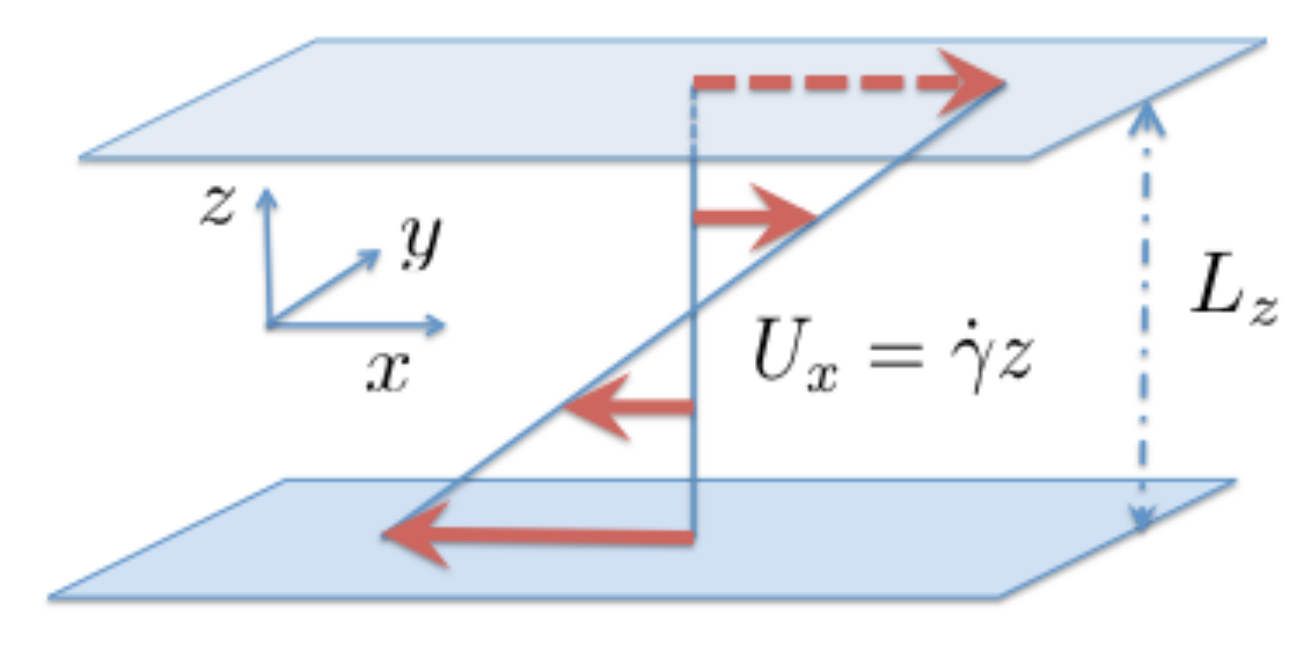}
\put(-112,65){(a)}
\hspace{0.3cm}
\includegraphics[width=0.35\linewidth]{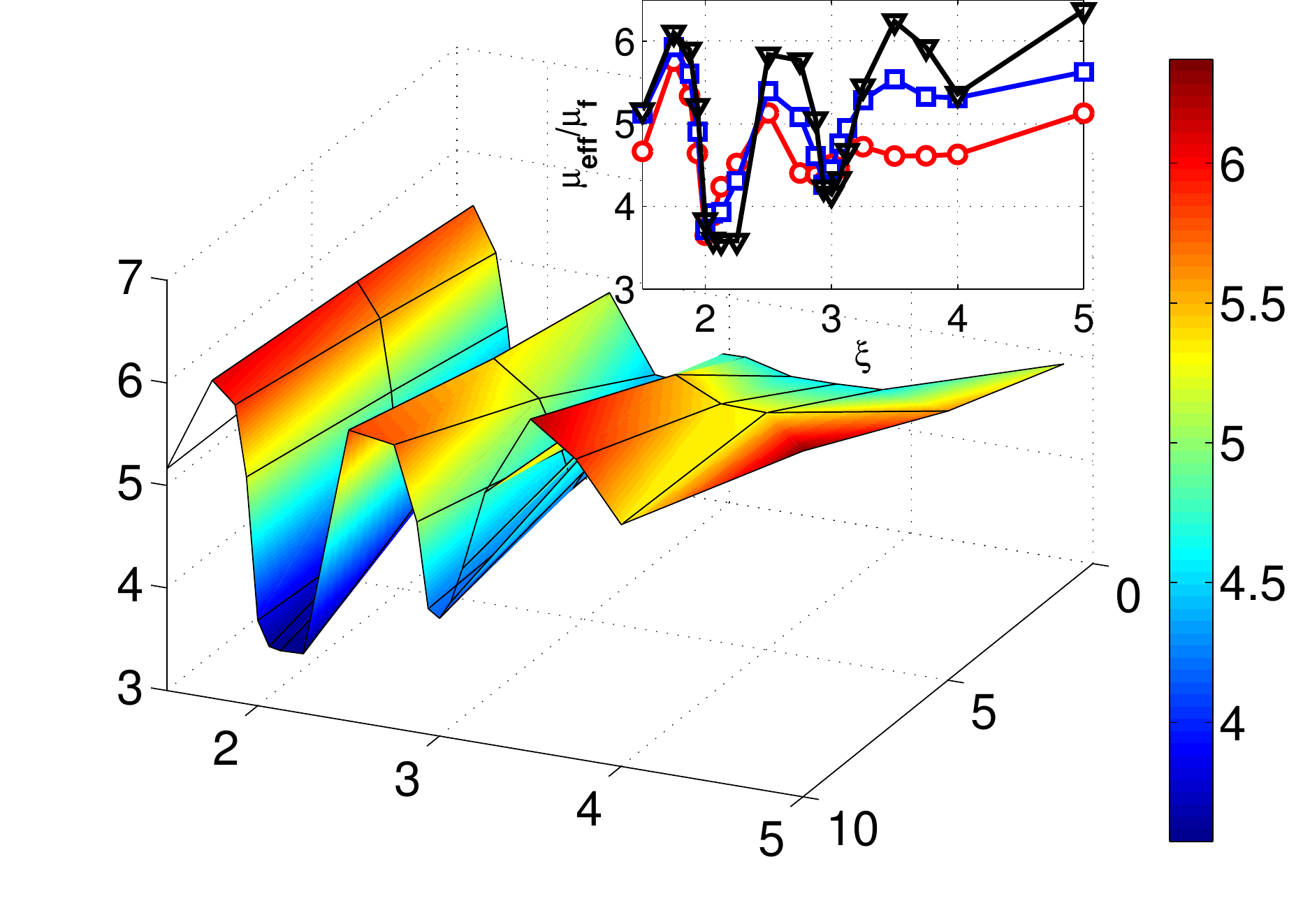}
\put(-150,90){ (b) }
\put(-112,-3){\large{$\xi$}}
\put(-42,15){$\Rey$}
\put(-174,40) {\rotatebox{90}{$\mue/\muf$} }
\includegraphics[width=0.32\linewidth]{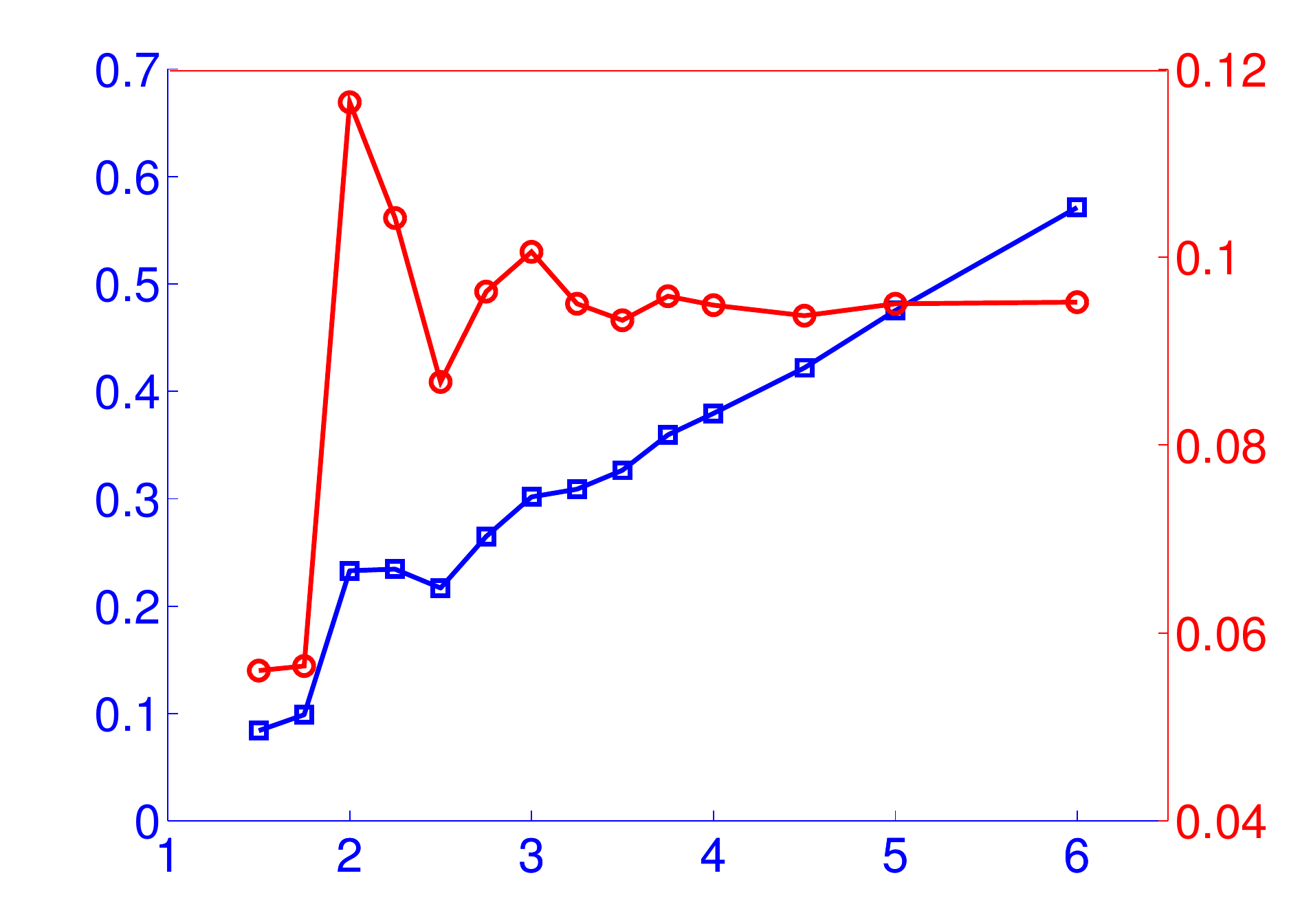}
\put(-80,95){(c)}
\put(-82,-5){  \large{$\xi$}  }
\put(-162,60){ \rotatebox{90}{ $\FF$ }          }
\put(-30,60){  \rotatebox{90}{ $\ff$ }    }
\caption{\label{fig:mue} (a) A sketch of the domain. 
(b) Surface plot of the effective viscosity $\mue/\muf$ as a function of $\xi$
and $\Rey$ for $\phi=0.3$. The inset shows $\mue/\muf$ versus $\xi$ 
for $\phi=0.3$ and for three different values of $\Rey=1$ (red circles), $5$ (blue squares) and 
$10$(black triangles).
(c) The flow rate of matter (both solvent and
additive ) $\FF$ (blue squares, left vertical axis), defined in \eq{eq:flux},
and the flow-rate-per-unit-cross-section, $f$, (red circles, right vertical axis)
as a function of $\xi$, for $\Rey=5$ and $\phi=0.3$. 
}\end{center}
\end{figure*}
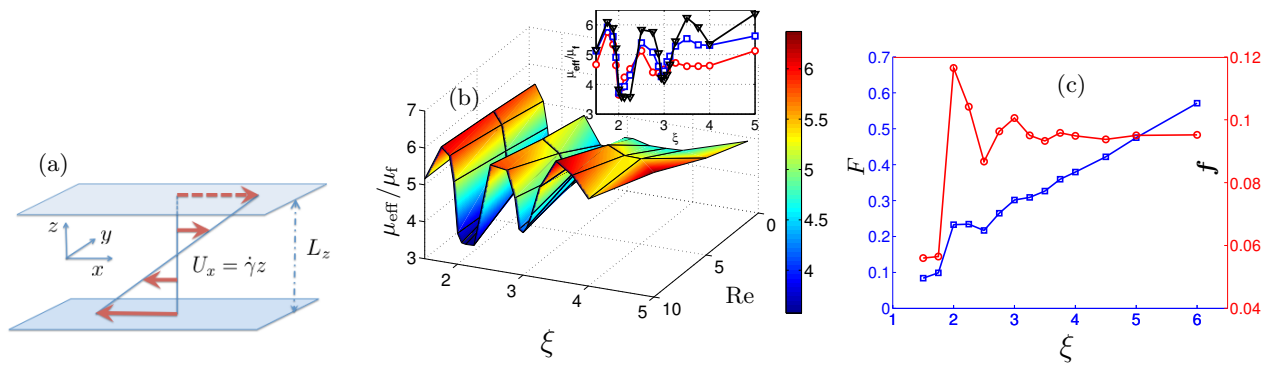

As a concrete example, we consider direct numerical simulations (DNS) 
of  three-dimensional plane-Couette flow -- with the
$x$, $y$ and $z$ coordinates along the stream-wise, span-wise and wall-normal 
directions respectively [see \subfig{fig:mue}{a}] -- embedded with neutrally-buoyant rigid spheres of radius $\a$. 
The fluid phase is described by solving the incompressible
Navier--Stokes equations in three dimensions. The fluid
is sheared by imposing constant stream-wise velocity of 
opposite signs, $\Uzero=\gdot \Lz/2 $ at $z=\pm\Lz/2$. 
Periodic boundary conditions are imposed on the other two
directions ($x$ and $y$ with lengths $\Lx$ and $\Ly$ 
respectively). The motion of the rigid spheres and their interaction with the
flow is fully resolved by using the Immersed Boundary Method~\cite{pes02,mit+iac05}.
A description of the equations and the details of the algorithm
is provided as supplemental material.

\begin{figure}[]
\begin{center}
\includegraphics[width=0.8\columnwidth]{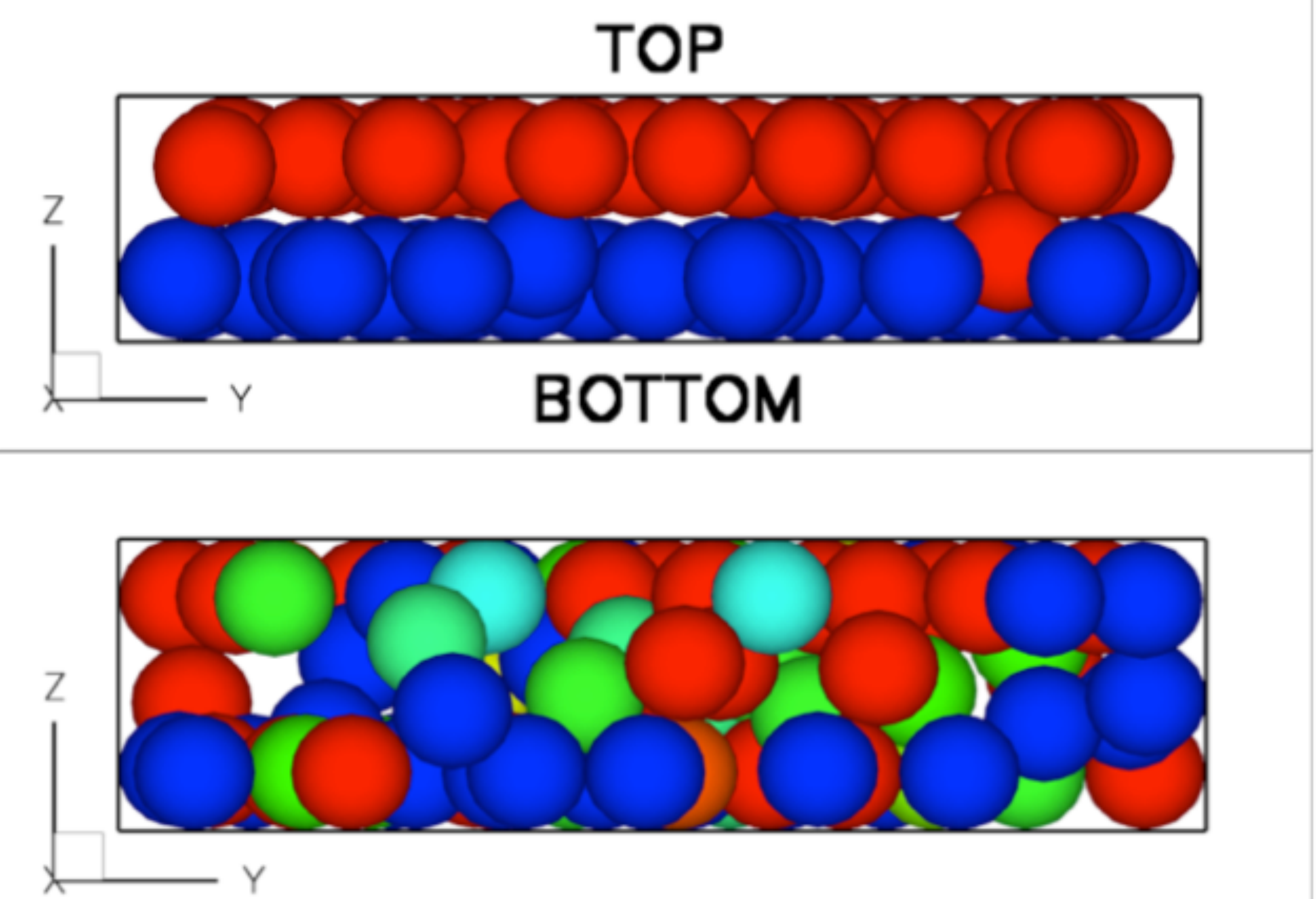}
\put(-25,65) { \large{ (a) }  } \\
\includegraphics[width=0.9\columnwidth]{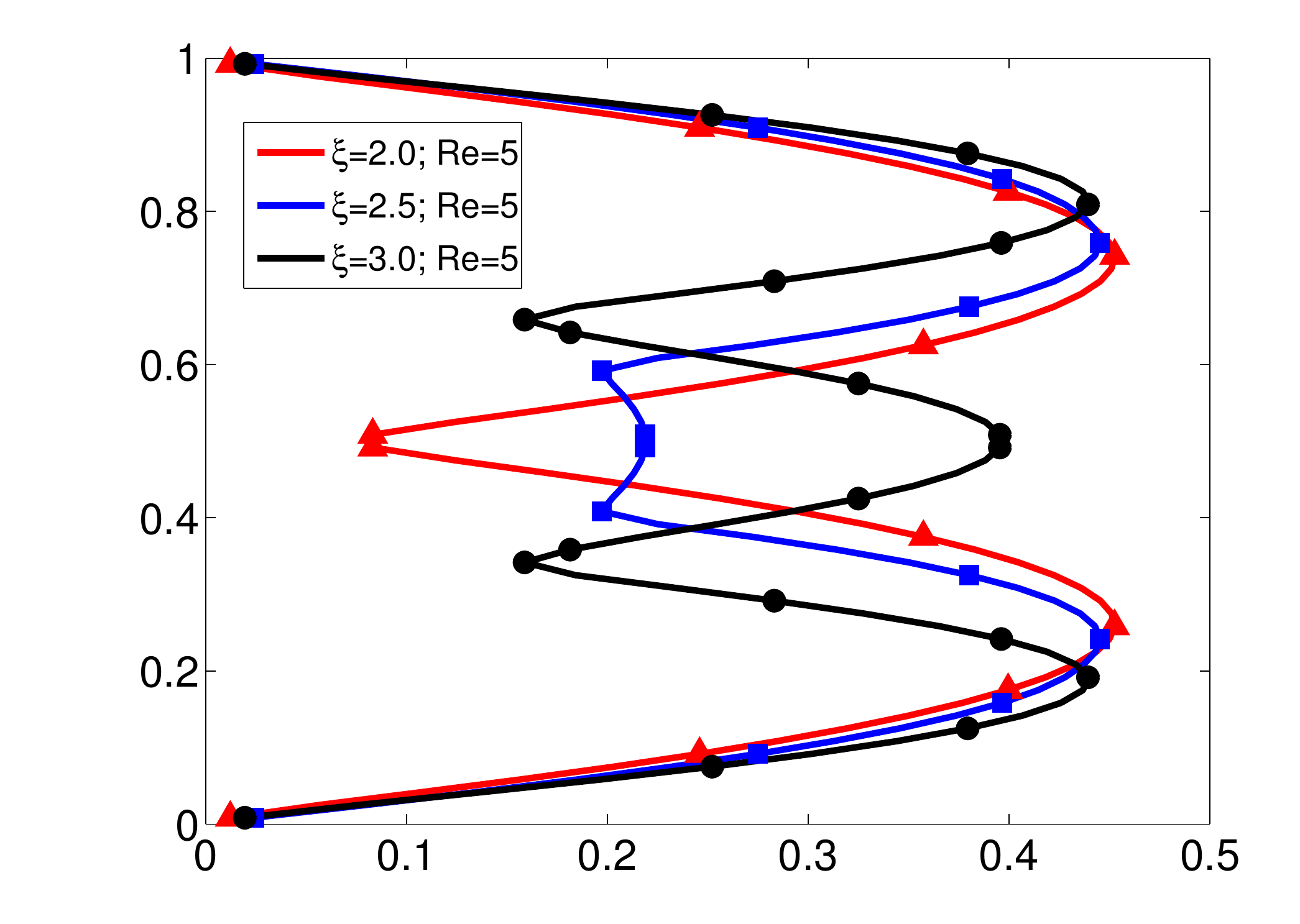}
\put(-50,80) { \large{ (b) } } 
\put(-120,-5){ \large{$\rhop$} }
\put(-210,60){\rotatebox{90}{ $z/\Lz$}}
\caption{\label{fig:layer} 
(a) Three dimensional view of positions of the spheres 
for $\Rey=5$, $\phi=0.3$ and  $\xi=2$ (top) and  $\xi=2.5$ (bottom) at later times after the initial transients. 
The particles are color coded by their initial wall-normal position. 
(red: close to the top boundary, blue: close to the bottom boundary).
(b) Average local density of spheres $\rhop$ versus the wall-normal coordinate 
$z$, for $\xi=2$ (red triangles), $2.5$ (blue squares), and $3$ (black dots). 
}\end{center}
\end{figure}

\begin{figure}[]
\begin{center}
\includegraphics[width=0.9\columnwidth]{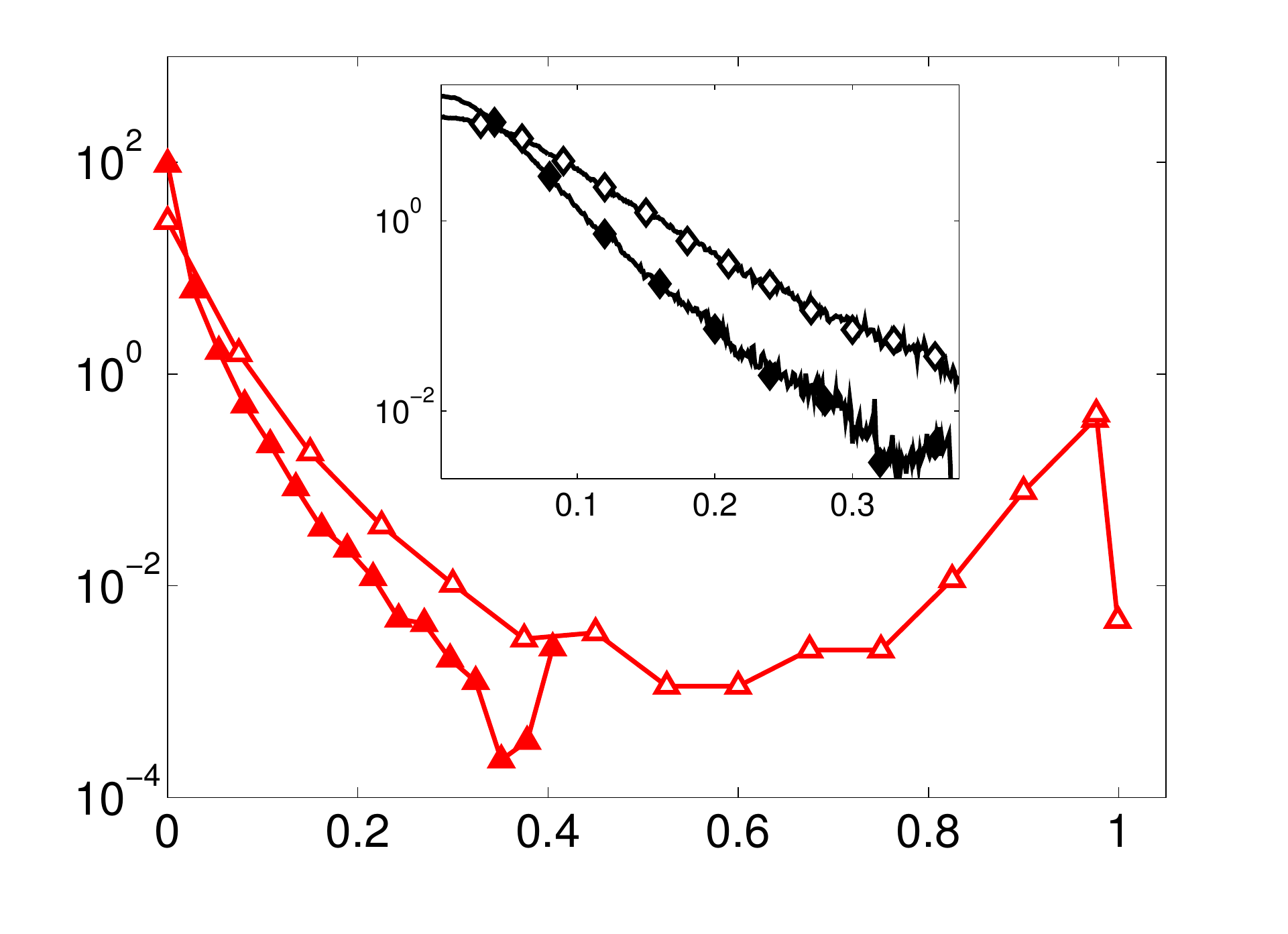}
\put(-50,120){ \large{ (a) }}
\put(-115,0){{ \large $\dz/\Lz$}}
\put(-225,70){\rotatebox{90}{ \large $P(\dz)$}}
\put(-100,60){ $\dy/\Ly$ }
\put(-170,90){\rotatebox{90}{  $P(\dy)$} }\\
\includegraphics[width=0.9\linewidth]{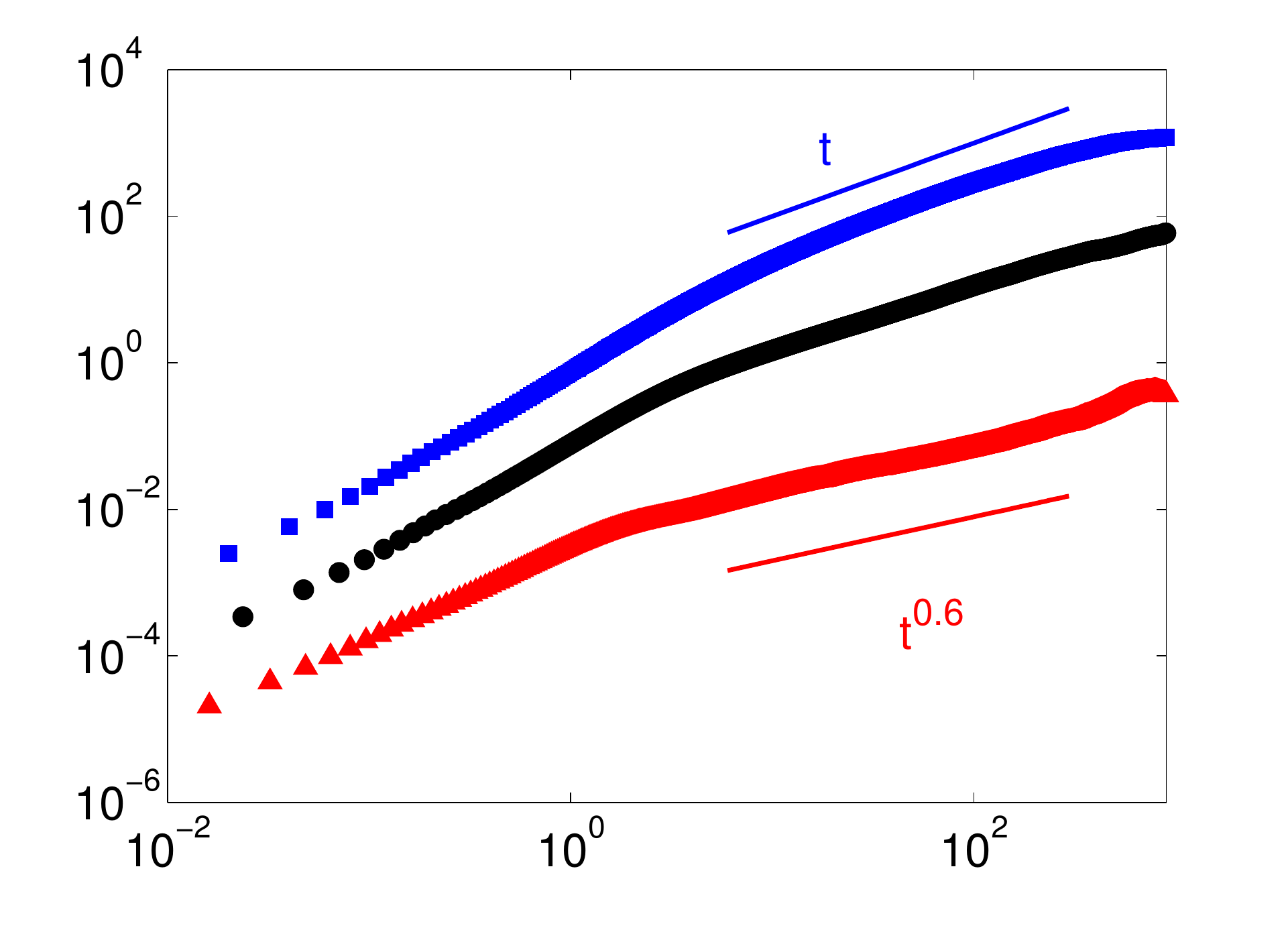}
\put(-180,120){ \large{ (b) }}
\put(-115,0){{ \large$\dot\gamma t$}}
\put(-225,70){\rotatebox{90}{ \large $\Sz$}}
\caption{\label{fig:PDFdz} 
(a) The PDF, $P(\dz)$, of the displacement ($\dz$) of the center of the spheres 
along the $z$ direction for $\xi=2$,  at early times, $t=2.5\igdot$ (red, filled triangles) 
and late times $t=386\igdot$ (red, open triangles). For comparison, the inset shows
the PDF, $P(\dy)$, of the displacement ($\dy$) 
along the $y$ direction at early times, $t=2.5\igdot$ (black, filled diamonds) 
and late times $t=5\igdot$ (black, open diamonds)
(b) Log-log plot of mean square displacement of the spheres, $\Sz$, versus $t\gdot$.
At late times,  $\Sz \sim t^{\beta}$,  where $\beta = 1$  implies diffusion.
For three different values of $\xi=2$ (red triangles), $2.5$ (blue squares) and $3$ (black dots), we obtain 
$\beta \approx 0.61$, $0.95$ and $0.88$,  respectively.
}\end{center}
\end{figure}

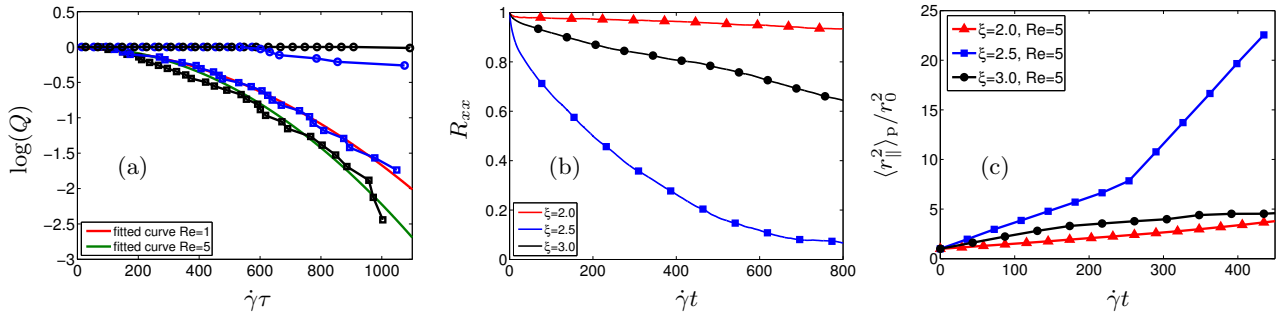
\begin{figure*}[]
\begin{center}
\includegraphics[width=0.32\linewidth]{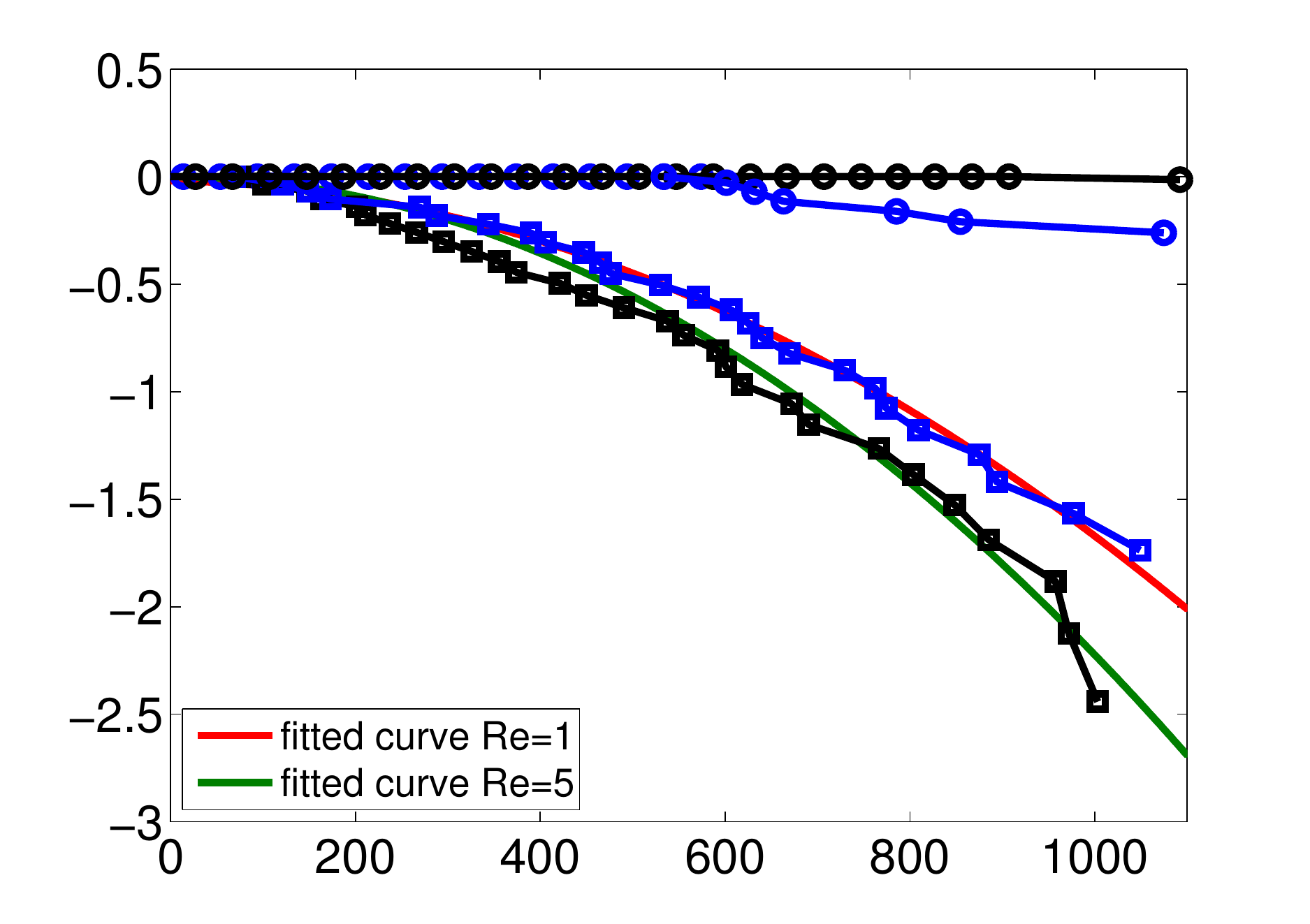}
\put(-130,45){ (a) }
\put(-168,45){\rotatebox{90}{$\log(Q)$} }
\put(-82,-5){ $\gdot \tau$ }
\includegraphics[width=0.32\linewidth]{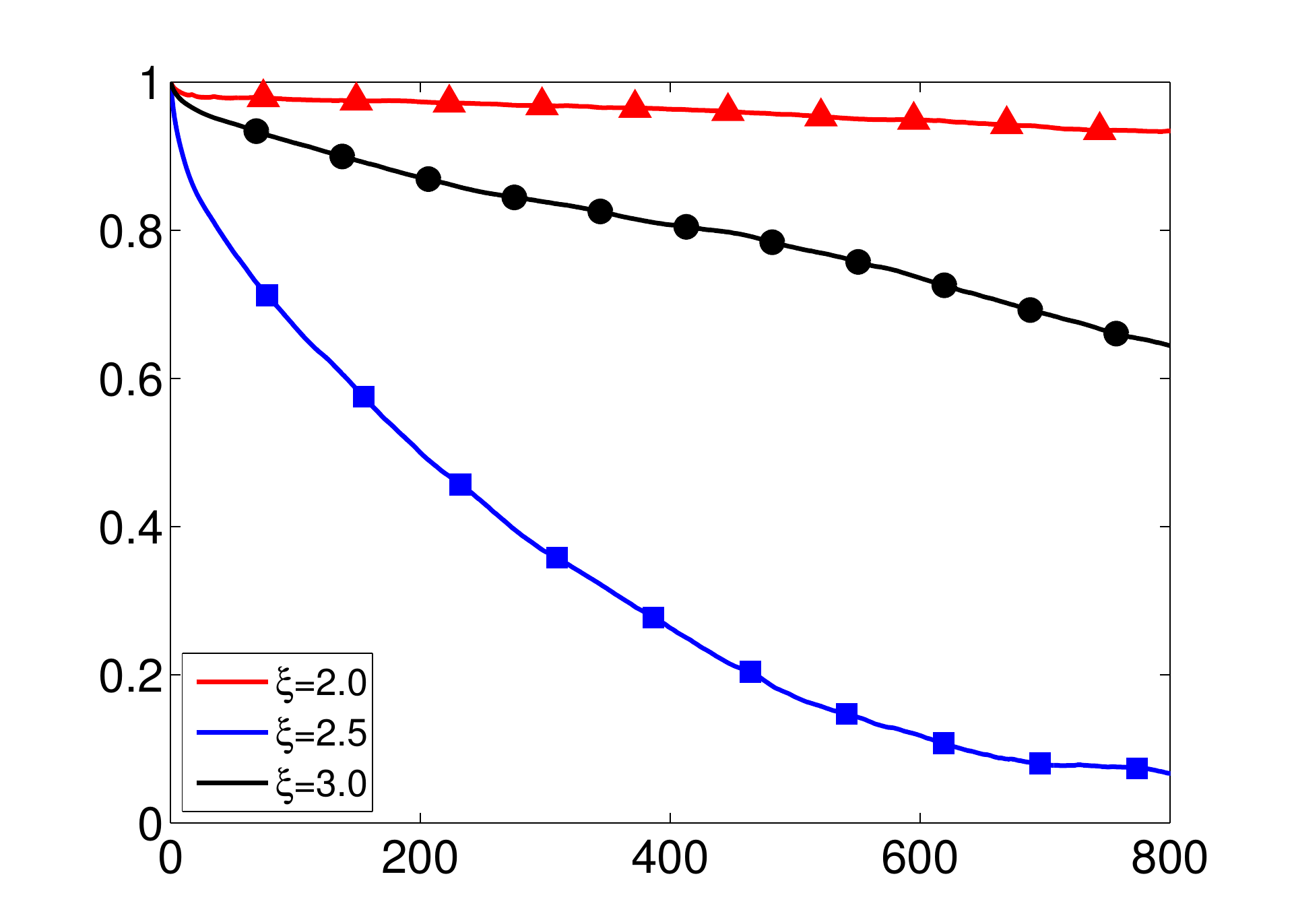}
\put(-130,45){ (b) }
\put(-164,55){\rotatebox{90}{ $R_{xx}$ } }
\put(-82,-5){ $\gdot t$ }
\includegraphics[width=0.32\linewidth]{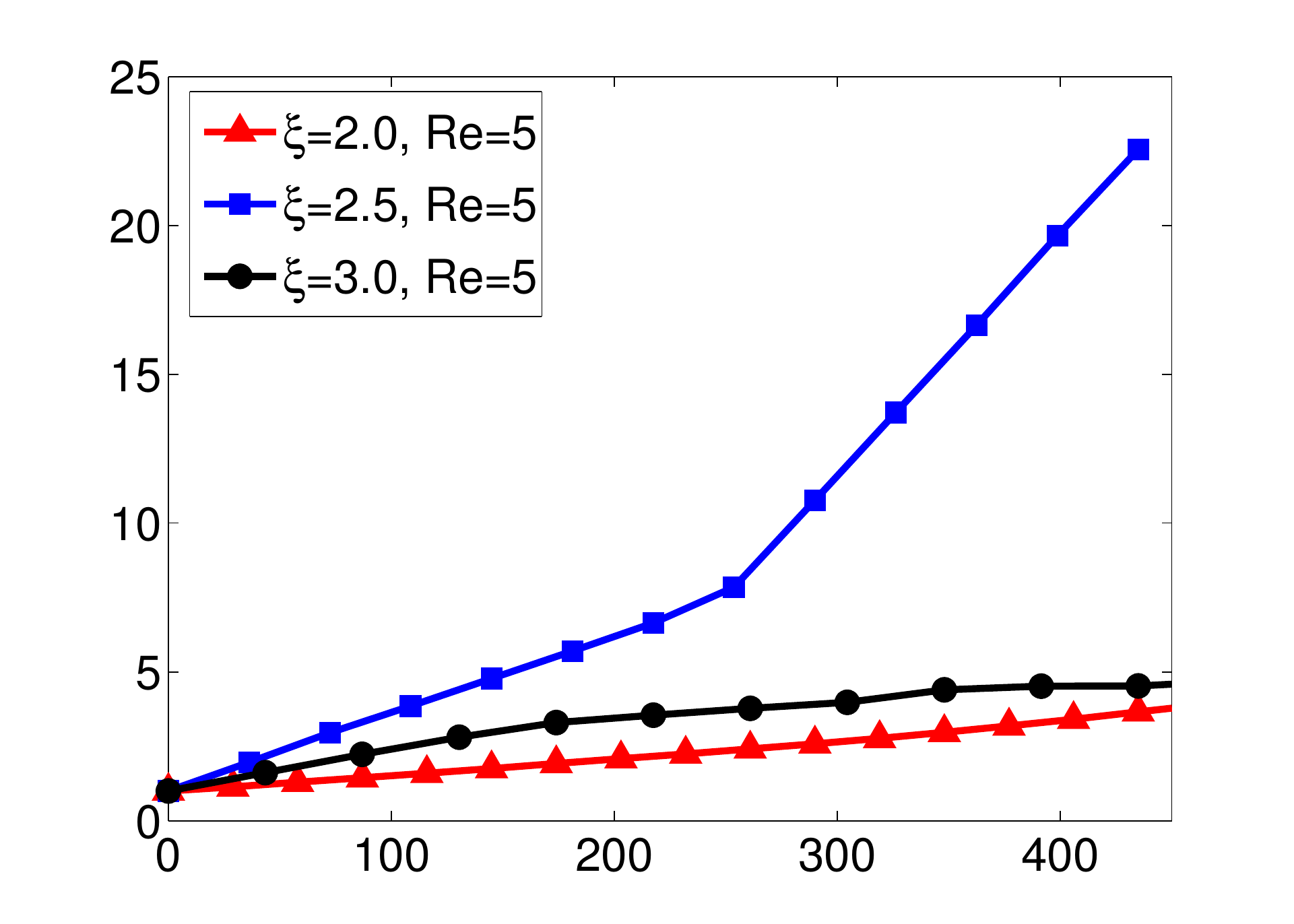}
\put(-130,45){ (c) }
\put(-168,40){\rotatebox{90}{  $\bra{\rparl^2}_{\rm p}/r_0^2$} }
\put(-82,-5){ $\gdot t$ }
\caption{\label{fig:frozen} 
(a) Cumulative PDF, $Q$, of the residence times of spheres ($\gdot\tau$)
in a layer for $\phi=0.3$, $\Rey=1$ (in blue) ,  and $\Rey=5$ (in
black) and   $\xi=2$ (circles), and $\xi=2.5$ (squares).
(b) The streamwise velocity-velocity auto-correlation function, $R_{xx}$, for $\Rey=5$ and
$\phi=0.3$ for four different values of $\xi=2$(red triangles), $2.5$ (blue squares), 
and $3$ (black dots).
(b) Evolution of averaged square of relative distance between pairs of sphere, $\bra{\rparl^2(t,\rzero)}_{\rm p}$ 
which were at distance $\rzero$ at $t=0$;
for $\rzero=4\a$, for three different values of 
$\xi=2$ (red triangles), $2.5$ (blue squares), and $3$ (black dots).
}
\end{center}
\end{figure*}

We study the effects of confinement by changing the dimensionless ratio
$\xi\equiv\Lz/2\a$, where $\Lz$ is the channel width  in the $z$ direction. 
In practice, we change $\Lz$ but hold the particle radius $\a$ 
fixed. The effective viscosity, $\mue$, is thus function of
the dimensionless numbers, $\phi$, $\xi$ and the particle Reynolds number,
$\Rey\equiv \rho\gdot\a^2/\muf$, where $\rho$ and $\muf$ are 
the density and dynamic viscosity of the solvent.  
The most striking result of our simulations is that at or near integer
values of $\xi$, $\mue$ decreases significantly compared to its bulk
value; see ~\subfig{fig:mue}{b}. 
This drop can be so large that the net rate of mass transport for
a thinner channel ($\xi=2$) is more than in a wider channel
($\xi=2.5$), see \subfig{fig:mue}{c}. 
We further demonstrate that, at small integer values of $\xi$, the 
rigid spheres self-organize into an integer number of horizontal layers, 
which slide on one another, see \fig{fig:layer}.
This, in turn, decreases the transport of horizontal 
momentum across the layers generating the drop in effective viscosity. 
We also show that at integer values  of $\xi$, for which layers appear: 
(a)  the movement of spheres across layers is {\it not} a diffusive
process, \fig{fig:PDFdz},
(b) the typical residence time of the spheres in a layer is much longer
than the time scale set by the shear, $\igdot$; in fact, 
a large number of spheres never leave the layer within the runtime of
the simulations~\subfig{fig:frozen}{a}; 
(c)  the layers are structurally liquid-like but dynamically 
very slow compared to the timescale set by $\igdot$, see
\fig{fig:frozen}. 

Initially, the spheres are placed at random locations, with no
overlap between each other, and with velocities that are equal to the local fluid velocity,
which is taken to be  the laminar Couette profile, \subfig{fig:mue}{a}.
We calculate the effective viscosity, $\mue$, as
the ratio between the tangential stress at the walls and the shear $\gdot$
averaged over the walls and over time. 
See supplemental material for  further details of measurement and estimate of errors.

The effective viscosity $\mue$ is shown in \fig{fig:mue}{b} as a function of  $\Rey$
and  $\xi$, for $\phi=0.3$. For large channel widths, e.g., $\xi=5$, and large volume-fractions,
e.g, $\phi\ge 0.2$, we obtain  shear-thickening~[see Supplemental Material \fig{fig:mue_phi}] 
as was found earlier in Ref.~\cite{pic+bre+mit+bra13}.
 
Here we address how confinement affects the rheology.
Experiments~\cite{pey+ver11}, in agreement with 
earlier analytical calculations~\cite{dav+pey08}, have found that at  small volume-fraction ($\phi \le
0.15$), $\mue$ increases as $\xi$  decreases. 
Our results [Supplemental Material, \fig{fig:mue_phi}] also supports this 
conclusion.  Furthermore, our simulations can access hitherto unstudied higher
values of volume-fraction ($\phi \ge 0.2$) for which this trend seems to reverse, i.e., $\mue$  decrease with confinement, 
see e.g., Supplemental Material, \fig{fig:mue_phi}. 
On deeper scrutiny, a more striking picture emerges. As we show in \subfig{fig:mue}{b}, for  $\phi=0.3$,
at or near small {\it integer} values of $\xi$ ($\xi=2,3$, and $4$), the effective viscosity 
drops significantly.  In our simulations, the minimum value of viscosity is found at $\xi=2$, 
which is  as low as $50\%$ of its bulk value  ($\xi \ge 6$). 

To appreciate how dramatic this effect is, we measure the 
(dimensionless) flux of matter (i.e. both the fluid and the spheres) through the channel, defined as ~\cite{pic+bre+bra15} -
\begin{equation}
\FF \equiv  \frac{1}{\gdot\Lz\a^2}\rho  \bra{\int \Uc_x p(z) dydz};
\label{eq:flux}
\end{equation} 
where $\Uc \equiv \zeta\Up + (1-\zeta)\uu$, with $\zeta=1$ at a grid point inside
a rigid sphere but $\zeta=0$ otherwise~\footnote{The phase indicator $\zeta$ is
is related to the volume fraction by $\phi = (1/V)\int \zeta dV $
where $V$ is the total volume of the channel.},
Here, $\Up$ is the velocity of the sphere,
$\uu$ is the velocity of the fluid, 
$p(z) = 1$ for $z\ge0$ and $-1$ for $z<0$, 
and $\bra{\cdot}$ denotes averaging over time.
In \subfig{fig:mue}{c} we report $\FF$  as a function of $\xi$.  
As the flux $\FF$ is not normalized by the cross-sectional area of the channel 
it is  expected to increase linearly as a function of  $\xi$.
This expectation indeed holds for $\xi > 4$. But, below that, for integer values of $\xi$
the effective viscosity can decrease so much that 
 $\FF$ is not even a monotonic function of $\xi$, in particular, 
$\FF (\xi=2) > \FF(\xi = 2.5)$; the flux through a wider channel is actually smaller. 
The flux-per-unit-area, $\ff \equiv \FF (a^2/L_yL_z)$, shown  in \subfig{fig:mue}{c} is expectedly a constant at large $\xi$. 
For  the small integer values of $\xi$, it is significantly higher than its bulk value.

To investigate the mechanism behind the rheology, 
we examine snapshots of the spheres, see \subfig{fig:layer}{a} for $\xi=2$ (top) and  $\xi=2.5$ (bottom). 
The spheres are color-coded by their initial wall-normal locations (red corresponds to an initial position 
near the top boundary and blue to the bottom boundary). 
It can be clearly observed, that for $\xi=2$, the particles form a bi-layered structure. 
This layering is also confirmed by the wall-normal profiles of the  average particle number density 
displayed in \subfig{fig:layer}{b}  for $\xi=2, 2.5$ and $3$. 
In the first and the last case, one can observe equally-spaced two and three prominent peaks respectively.
Note that a weaker layering is observed for $\xi=2.5$. 
The drop in effective viscosity thus corresponds to the formation of layers
that slide on each other, with little transport of momentum across the layers. 
For $\xi=2$, where layering is most prominent, 
the particles form disordered liquid-like structures, within each layer, as seen by the radial distribution function
of the position of the spheres, see Supplemental Material, \fig{fig:rdf}.

In order to understand the dynamics of particles in the wall-normal direction, 
we show in \subfig{fig:PDFdz}{a}, the probability distribution function (PDF) of the 
displacement of the spheres, $\dz(t) \equiv \zc(t) - \zc(0)$, at different times,
where $\zc(t)$ is the the $z$ coordinate of the center of a sphere.
Obviously, as  $t\to 0$ the PDF [$P(\dz)$] must approach a Dirac delta function.
Remarkably, for  $\xi=2$  the PDF has exponential tails, i.e., it is non-diffusive, with some
particles undergoing larger displacements as shown in \subfig{fig:PDFdz}{a}.
At later times, for $\xi=2$, the PDF of $\dz$ develops a 
peak at $d_z/L_z=1$, indicating the hopping between two layers.  
A similar behavior is observed for $\xi=3$ too.
Contrast this result with the PDF of the displacement along the span-wise($y$) direction, 
inset of \subfig{fig:PDFdz}{a}, which clearly shows Gaussian behavior at all times, implying diffusive dynamics. 
The second moment of these PDFs provides the mean squared displacement of the particles,
$\Sz(t) = \bra{ [\zc(t) - \zc(0)]^2}_{\rm p}$, the time evolution of which is shown in  \subfig{fig:PDFdz}{b}.
At late times, in general, a power-law dependence on time is found, $\Sz(t) \sim t^{\beta}$, where
$\beta = 1$ would imply a simple diffusive behavior.
For $\xi=2,3$ we estimate $\beta\approx 0.61$ and $0.88$ respectively. 
But for $\xi=2.5$ , $\beta\approx 1$ is obtained. 
The diffusive behaviour for $\xi=2.5$ can be  further confirmed by plotting the PDF of $\dz$. 
At intermediate times ($t=2.5\dot\gamma^{-1}$), for $\xi=2.5$, the PDF develops  Gaussian tails, 
indicative of a diffusive process and at late times it approaches a constant (see Supplemental Material,
\fig{fig:pdf}).

To visualize the dynamics, we provide movies of the particles' trajectories~(available at 
\url{https://www.youtube.com/watch?v=Qn4DiXZFsbw} for the
case $\xi=2$ and at 
\url{https://www.youtube.com/watch?v=AmNsAsY0eC8} for the 
case $\xi=2.5$ ). These clearly demonstrate that for  $\xi=2$,
each horizontal layer is structurally disordered but dynamically frozen. 
We quantify this phenomenon by three different measurements:

(A) We calculate the residence time ($\tau$) of a sphere in a single layer. 
A sphere is considered to reside within a horizontal layer till the wall-normal 
coordinate of its center, $\zc$, is within a distance of $2\a$ of its initial position.
Instead of calculating the PDF by histograms of $\tau$, we calculate the cumulative 
PDF, $Q(\tau)$, by the  rank-order method~\cite{mit+bec+pan+fri05}, as the 
latter is free from binning errors.
The cumulative PDF, $Q(\tau)$, is displayed in \subfig{fig:frozen}{a} as a function of $\tau$
for $\xi=2$ and $2.5$ for $\Rey=1$ and $5$.
For $\xi=2$, $Q$ remains very close to unity during the whole duration of the simulation,
i.e., very few spheres actually move from one layer to another. 
In other words, the layers are quite stable to perturbations in wall-normal directions. 
Conversely for $\xi=2.5$, $Q(\tau)$ can be fitted by a Gaussian.  

(B) The streamwise velocity  auto-correlation function of the spheres, 
$R_{xx} \equiv \langle U^{\rm p}_x(t)U^{\rm p}_x(0) \rangle$, 
is shown in \fig{fig:frozen}{b}. 
For $\xi=2$,  $R_{xx} \approx 1 $ 
 for a very long time, implying that the temporal fluctuations of the stream-wise velocity are
negligible. This suggests that each sphere moves in a layer with a uniform stream-wise
velocity keeping their relative distances practically constant.  For the cases
where the layering is not very strong, e.g., $\xi=2.5$ the auto-correlation function
decays rapidly.  For $\xi=3$, layering reappears and $R_{xx}$ again shows slow decay
in time. 

(C) Let us define $\rparl(t;\rzero) $ to be the (horizontal) distance between a pair of spheres 
at time $t$, which were at a (horizontal) distance $\rzero$ and $t=0$.  
If the layers formed by the spheres were truly frozen-in-time
we would obtain $\rparl(t;\rzero) = \rzero $ for all $t$ and $\rzero$. 
In \subfig{fig:frozen}{c}, we show the time evolution of  $\bra{\rparl^2(t;\rzero)}_{\rm p}$,
for $\rzero = 4\a$, where the symbol $\bra{\cdot}_{\rm p}$ denotes averaging over 
all possible particle pairs~\footnote{To take into account the movement of the spheres across layers,
we calculate $\rparl$  up to the time it takes for the wall-normal distance between the
pair of spheres to become larger than $2\a$. }.
 Had the spheres moved chaotically within a layer, $\rparl$ would have 
grown exponentially with time. Clearly for all the cases shown in \subfig{fig:frozen}{c},
$\rparl$ grows at most linearly with time. In particular, when layering occurs
($\xi=2,3$), it takes a long time ($t\ge 430\igdot$) for $\rparl(t;4\a)$ to grow by 
a factor of two. This quantifies again the dynamical stability of the layer, the relative
in-plane distance between pairs of spheres change slowly (linearly) with time. 

%

In conclusion, using numerical simulations, we demonstrate that the effective viscosity
of an extremely confined non-Brownian suspension  can exhibit a non-monotonic dependence on
the channel width, in particular the effective viscosity sharply decreases if the channel width is 
an (small) integer multiple of particle diameter. 
We demonstrate that this  behavior is accompanied by a change in micro-structure, namely the formation of 
particle layers parallel to the confining plates. 
The two-dimensional layers formed by the particles slide on each other, the layers are structurally liquid-like
by evolve on very show time scales. 
Similar layering under shear, have been theoretically anticipated \cite{zurita-gotor12}, but the
consequences for transport in extreme confinement as shown by us has never been demonstrated before.
 We finally note that our results are in contrast with the case of sheared Brownian suspensions
where the structure was found to be uncorrelated with measured viscosity \cite{xu+rice+dinner12}.

\section{Acknowledgment}
LB and WF acknowledge financial support by the European Research
Council Grant No. ERC-2013-CoG-616186, TRITOS and computer time
provided by SNIC (Swedish National Infrastructure for
Computing). DM acknowledges financial support from Swedish Research Council under grant 
2011-542 (DM) . PC thanks NORDITA for hospitality.

\newpage

\appendix*
\section{Supplemental Material}
Here we provide supplemental material for our paper. The material is
divided into several subsections. 
\subsection{Direct numerical simulation}
Simulations have been performed using the numerical code originally described in Ref.~\cite{bre12} that fully
describes the coupling between the solid and fluid phases. The Eulerian fluid phase is evolved according to the incompressible
Navier-Stokes equations,
\begin{equation}
\label{div_f}
\dive \uu = 0
\end{equation}
\begin{equation}
\label{NS_f}
\delt  \uu + \uu \cdot \grad \uu = -\frac{1}{\rho}\grad p + \nu \lap \uu + \ff
\end{equation}
where $\uu$, $\rho$ and $\nu=\mu/\rho$ are the fluid velocity, density and kinematic viscosity respectively ($\mu$ is
the dynamic viscosity), while $p$ and $\ff$ are the pressure and a generic force field (used to model the presence of particles).
The particles centroid linear and angular velocities, $\Up$ and $\Op$ are instead governed by the Newton-Euler Lagrangian
equations,
\begin{eqnarray}
\label{lin-vel}
\rhop \Vp \ddt{\Up} &=& \rho \oint_{\partial \mathcal{V}_{\rm p}}^{}   \bm{\taub} \cdot \nn\, dS\\
\label{ang-vel}
\IIp \ddt{\Op} &=& \rho \oint_{\partial \mathcal{V}_{\rm p}}^{} \rr \times  \bm{\taub} \cdot \nn\, dS
\end{eqnarray}
where $\Vp = 4\pi \a^3/3$ and $\IIp=2 \rhop \Vp \a^2/5$ are the particle volume and moment of inertia; $\bm{\taub} = -p \bm{ I} + 2\mu \EE$
is the fluid stress, with $\EE = \left(\grad \uu + \grad  \uu^{\rm T} \right)/2$ the deformation tensor; $\rr$ is the distance vector
from the center of the sphere while $\nn$ is the unity vector normal to the particle surface $\partial \mathcal{V}_p$. Dirichlet boundary
conditions for the fluid phase are enforced on the particle surfaces as 
$\uu|_{\partial \mathcal{V}_p} = \Up + \Op \times \rr$.\\
An immersed boundary method is used to couple the fluid and solid phases. The boundary
condition at the moving particle surface is modeled by adding a force field on the right-hand side 
of the Navier-Stokes equations. The fluid phase is therefore evolved in the whole
computational domain using a second order finite difference scheme on a staggered mesh while the time integration is performed by a third
order Runge-Kutta scheme combined with a pressure-correction method at each sub-step. This integration scheme is also used for the Lagrangian
evolution of eqs.~(\ref{lin-vel}) and (\ref{ang-vel}). Rach particle surface is described by uniformly distributed $N_{\rm L}$ Lagrangian points.
The force exchanged by the fluid on the particles is imposed on each $l-th$ Lagrangian point. This force is related to the Eulerian force field
$\ff$ by the expression $\ff(\xx) = \sum_{l=1}^{N_L} \FF_{\rm l} \delta^3(\xx - \Xl) \Delta V_l$ (where $\Delta V_{\rm l}$
is the volume of the cell containing the $l-th$ Lagrangian point while $\delta^3$ is the three dimensional Dirac delta distribution). 
The force field is calculated
through an iterative algorithm that ensures a second order global accuracy in space. In order to maintain accuracy, eqs.~(\ref{lin-vel}) and
~(\ref{ang-vel}) are rearranged in terms of the IBM force field,
\begin{align}
\label{lin-vel-ibm}
\rhop \Vp \ddt{\Up} &= -\rho \sum_{l=1}^{N_l} \FF_{\rm l} \Delta V_{\rm l} + \rho \frac{d}{dt} \int_{\mathcal{V}_{\rm p}}^{} \uu \, dV \\
\label{ang-vel-ibm}
\IIp \ddt{\Op} &= -\rho \sum_{l=1}^{N_l} \rr_{\rm l} \times \FF_{\rm l} \Delta V_{\rm l} + \rho \frac{d}{dt} \int_{\mathcal{V}_{\rm p}}^{} \rr \times \uu\, dV 
\end{align}
where $\rr_{\rm l}$ is the distance from the center of a particle. The second terms on the right-hand sides are corrections to account for
the inertia of the fictitious fluid contained inside the particles. Lubrication models based on Brenner's asymptotic solution 
gare used to correctly reproduce the interaction between two particles when the gap
distance between the two is smaller than twice the mesh size. A soft-sphere collision model is used to account for collisions 
between particles and between particles and walls. An almost elastic rebound is ensured with a restitution coefficient set at 
$0.97$. These lubrication and collision forces are added to the right-hand side of eq.~(\ref{lin-vel-ibm}).\\
The code has been validated against several classic test cases~\cite{bre12}
and has been used earlier to study shear-thickening in
inertial suspensions in Ref.~\cite{pic+bre+mit+bra13}.
\subsection{Measurement of effective viscosity}
\begin{figure}[]
\begin{center}
\includegraphics[width=0.9\linewidth]{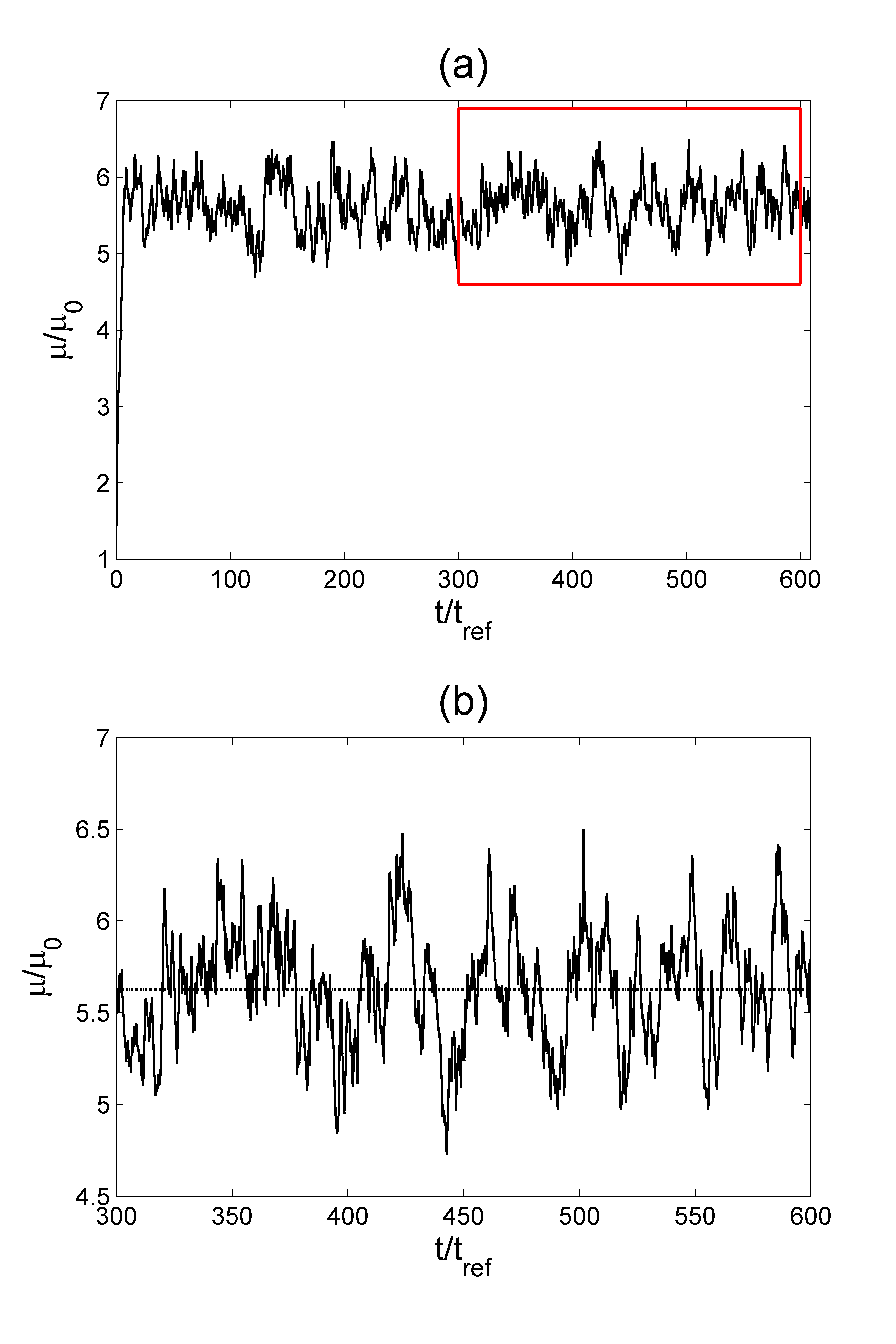}
\caption{\label{fig:trans} The effective viscosity
  $\mue(t)$ (normalized by the viscosity of the solvent) as a 
function of time for one representative run.}
\end{center}
\end{figure}

We calculate the effective viscosity  
as the ratio between the  tangential stress at the walls and $\gdot$.
The tangential stress, and consequently the effective viscosity, is different at different points of the wall and
also at each point changes as a function of time. At each time, we
calculate the average of this effective viscosity over the walls and
obtain a time-series $\mue(t)$.
This time-series is displayed in
\fig{fig:trans} for one particular case. Clearly, after a short while the effective viscosity reaches
a stationary value with fluctuations about it. 
Typically, we find that our simulations reach statistically 
stationary state when time  $t \ge 200 \gdot^{-1}$ 
The average of $\mue(t)$ over this statistically stationary state is the 
effective viscosity $\mue$ and the standard deviation of the
fluctuations is used as an estimate of the error in the measurement of
$\mue$ as reported in Table\ref{tab:datamu}
A careful look at the table will convince the reader that the relative strength of
the fluctuations decreases at commensuration. 
\begin{table}
  \begin{center}
  \begin{tabular}{cccc}
      $\xi$     & $\mue/\muf\,(\Rey=1)$   &   $\mue/\muf\,(\Rey=5)$ & $\mue/\muf\,(\Rey=10)$ \\[3pt]
      $1.5000$  & $4.665 \pm 0.113$           & $5.123 \pm 0.079$           & $5.161 \pm 0.107$  \\
      $1.7500$  & $5.758 \pm 0.116$           & $5.925 \pm 0.114$           & $6.100 \pm 0.115$  \\
      $1.8750$  & $5.334 \pm 0.186$           & $5.603 \pm 0.190$           & $5.895 \pm 0.173$  \\
      $1.9375$  & $4.641 \pm 0.190$           & $4.906 \pm 0.234$           & $5.213 \pm 0.209$  \\
      $2.0000$  & $3.650 \pm 0.120$           & $3.709 \pm 0.058$           & $3.831 \pm 0.066$  \\
      $2.0625$  & $3.888 \pm 0.152$           & $3.916 \pm 0.143$           & $3.596 \pm 0.075$  \\
      $2.1250$  & $4.237 \pm 0.183$           & $3.928 \pm 0.139$           & $3.574 \pm 0.100$  \\
      $2.2500$  & $4.517 \pm 0.141$           & $4.308 \pm 0.244$           & $3.581 \pm 0.087$  \\
      $2.5000$  & $5.129 \pm 0.165$           & $5.396 \pm 0.189$           & $5.836 \pm 0.181$  \\
      $2.7500$  & $4.403 \pm 0.162$           & $5.083 \pm 0.163$           & $5.766 \pm 0.158$  \\
      $2.8750$  & $4.381 \pm 0.147$           & $4.609 \pm 0.138$           & $5.051 \pm 0.148$  \\
      $2.9375$  & $4.303 \pm 0.147$           & $4.261 \pm 0.120$           & $4.223 \pm 0.102$  \\
      $3.0000$  & $4.506 \pm 0.155$           & $4.446 \pm 0.130$           & $4.148 \pm 0.180$  \\
      $3.0625$  & $4.462 \pm 0.177$           & $4.763 \pm 0.148$           & $4.326 \pm 0.223$  \\
      $3.1250$  & $4.718 \pm 0.185$           & $4.942 \pm 0.171$           & $4.670 \pm 0.237$  \\
      $3.2500$  & $4.721 \pm 0.184$           & $5.286 \pm 0.160$           & $5.449 \pm 0.172$  \\
      $3.5000$  & $4.608 \pm 0.165$           & $5.535 \pm 0.166$           & $6.238 \pm 0.189$  \\
      $3.7500$  & $4.610 \pm 0.168$           & $5.326 \pm 0.142$           & $5.923 \pm 0.166$  \\
      $4.0000$  & $4.629 \pm 0.156$           & $5.323 \pm 0.154$           & $5.358 \pm 0.176$  \\
      $5.0000$  & $5.150 \pm 0.208$           & $5.626 \pm 0.152$           & $6.371 \pm 0.193$  \\
      $6.0000$  & $4.752 \pm 0.180$           & $5.816 \pm 0.143$           & $6.807 \pm 0.173$  \\

  \end{tabular}
  \caption{Values of $\mue/\muf$ obtained at different $\xi$ and $\Rey$ with respective error.}
  \label{tab:datamu}
  \end{center}
\end{table}
\subsection{Effective viscosity as a function of volume fraction}
\label{sec:phi}
\begin{figure}
\begin{center}
\includegraphics[width=0.9\columnwidth]{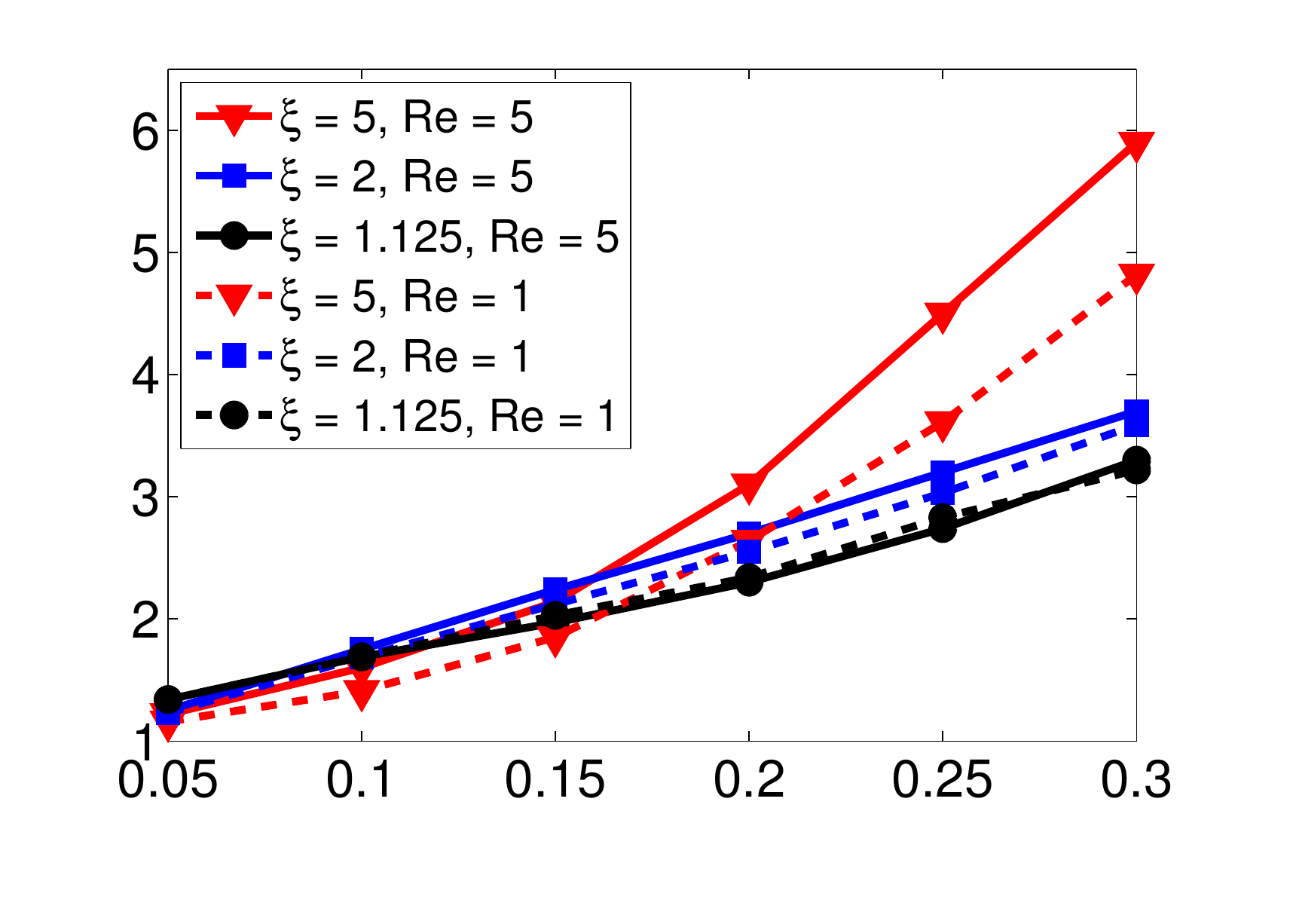}
\put(-110,2){{\large$\phi$}}
\put(-220,50) {\rotatebox{90}{ $\mue/\muf$}}
\caption{\label{fig:mue_phi} The effective viscosity $\mue/\muf$ versus $\phi$ for 
two different values of  $\Rey=1$ (dashed lines) and $5$ (continuous lines) for three different
values of $\xi=1$ (black filled circles), $2$ (blue squares) and $5$ (red triangles). 
}\end{center}
\end{figure}

Here we show,  see~\fig{fig:mue_phi}, the variation of the effective
viscosity, $\mue$ as a function of the volume fraction, $\phi$. 
\subsection{In-layer radial distribution function}
\label{sec:rdfi}
\begin{figure}[]
\begin{center}
\includegraphics[width=0.9\linewidth]{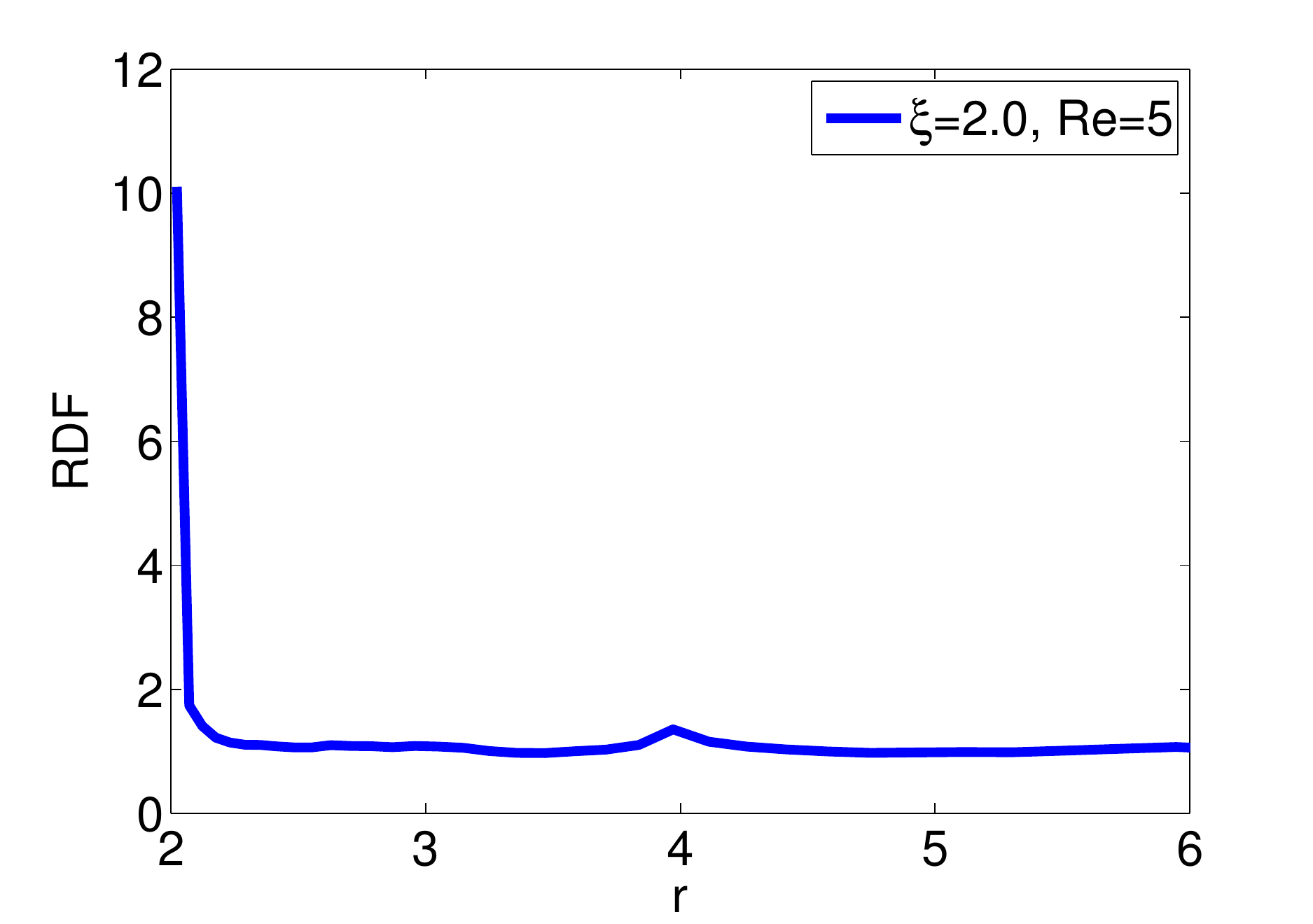}
\caption{\label{fig:rdf} 
The radial distribution function of the position of the 
centers of the spheres in the $x-y$ plane. 
}\end{center}
\end{figure}

The radial distribution function of the position of the 
centers of the spheres in the $x-y$ plane
is shown in \fig{fig:rdf}. This is defined as 
\begin{equation}
\Rparl \equiv \frac{1}{2 \pi r \Delta z n_0} \frac{dN_r}{dr}
\end{equation}
where $N_r$ is the number of particles is the number of particles in a cylinder of radius $r$ and 
$n_0=N_p(N_p-1)/(2V)$ is the density of particles pairs in a volume $V$ with 
$N_p$ the total number of particles. 
\begin{figure}[]
\begin{center}
\includegraphics[width=0.9\linewidth]{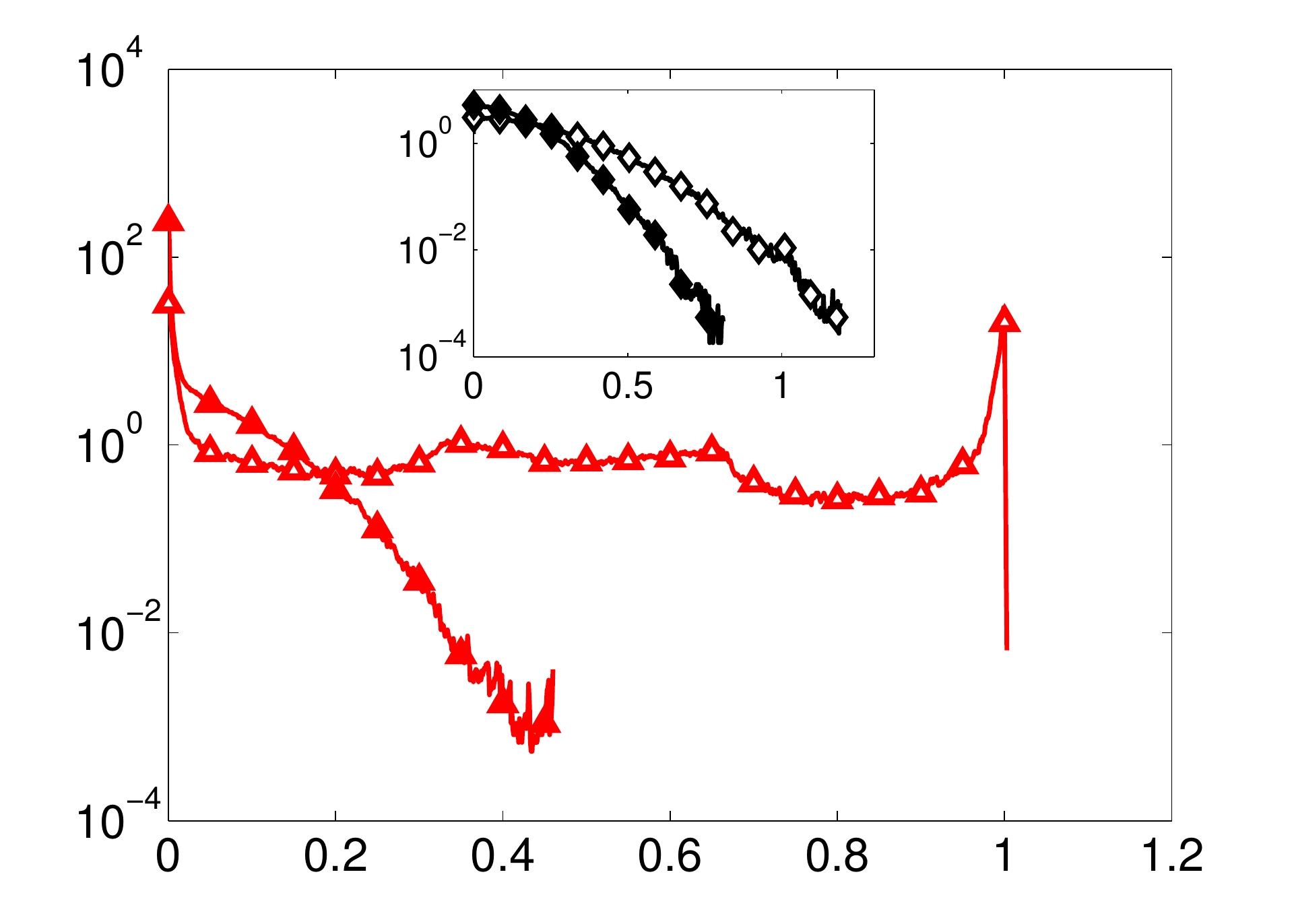}
\put(-115,0){{ \large $\dz/\Lz$}}
\put(-225,70){\rotatebox{90}{ \large $P(\dz)$}}
\put(-132,100){ $\dy/\Ly$ }
\put(-170,95){\rotatebox{90}{  $P(\dy)$} }\\
\caption{\label{fig:pdf} 
 The PDF ($P(\dz)$) of the displacement ($\dz$) of the center of the spheres 
along the $z$ direction, for $\xi=2.5$, at early times, $t=2.5\igdot$ (red, filled triangles) 
and late times $t=386\igdot$ (red, open triangles). For comparison, the inset shows
the PDF ($P(\dy)$) of the displacement ($\dy$) 
along the $y$ direction at early times, $t=2.5\igdot$ (black, filled diamonds) 
and late times $t=5\igdot$ (black, open diamonds)
}\end{center}
\end{figure}
\end{document}